\documentclass{article}
\usepackage{emulateapj}
\usepackage{float,epsfig,psfig}

\newcommand{\msun}{M$_{\odot}$}
\newcommand{\ergl}{ergs~s$^{-1}$}
\newcommand{\ergcms}{ergs~cm$^{-2}$~s$^{-1}$}
\newcommand{\DTF}{{$D_{25}$}}
\newcommand{\fdtf}{$f_{{\rm D}25}$}
\newcommand{\LCG}{L\'{o}pez-Corredoira \& Guti\'{e}rrez}
\newcommand{\ros}{{\sl RoSat}}
\newcommand{\etal}{et al.}
\slugcomment{Submitted to Astrophysical Journal Letters}

\begin{document}

\title{The Lack of Halo Ultraluminous X-ray Sources}

\author{
Douglas~A.~Swartz\altaffilmark{1}}
\altaffiltext{1}{Universities Space Research Association,
    NASA Marshall Space Flight Center, VP62, Huntsville, AL, USA}

\begin{abstract}

The premise that Ultraluminous X-ray sources (ULXs)
 exist beyond the optical extent of nearby galaxies is investigated.
A published catalog containing 41 ULX candidates located 
 between 1 and $\sim$3 times the 
 standard \DTF\ isophotal radius of their putative host galaxies is examined.
Twenty-one of these sources have spectroscopically-confirmed distances.
All 21 are background objects giving a 95\% probability that at least 37
 of the 41 candidates are background sources.
Thirty-nine of the 41 sources have X-ray-to-optical flux ratios,
 $-1.6<\log(F_{\rm X}/F_{\rm O})<+1.3$,
 consistent with those of background active galactic nuclei.
(The remaining two are not detected in optical to a weak limit of
 $m_B \sim 21.5$~mag
 corresponding to $\log(F_{\rm X}/F_{\rm O})\gtrsim 1.6$.)
The uniform spatial distribution of the sample is also consistent with 
 a background population.
This evidence suggests
 that ULXs rarely, if at all, exist
 beyond the distribution of luminous matter in nearby 
 galaxies and, as a consequence, there is no correlation  
 between the population of ULXs
 and halo objects such as old globular clusters or Pop~III remnants.

\end{abstract}


\section{Introduction}   

Ultraluminous X-ray sources (ULXs) are point-like
 non-nuclear sources in nearby galaxies with apparent X-ray luminosities 
 $L_{\rm X} \gtrsim 10^{39}$~\ergl.
Their nature is still a mystery:
if they are accreting sources at the distance of their putative host galaxies,
 then their high luminosities require either beamed emission geometries,
 or super-Eddington emission rates, or accretion onto compact
objects more massive than predicted by stellar evolution models.
See Fabbiano (2006) for a review of ULX theory, observation, and their importance
 in studies of extragalactic X-ray source populations.

Distances to candidate ULXs can, in principal, be determined 
 spectroscopically if they have bright optical counterparts.
Because spectroscopy requires high contrast between the optical counterpart
 and the underlying galaxy light,
 most spectroscopic measurements made to date have been of bright counterparts
 located in the low surface brightness regions at relatively 
 large distances from the centers of their putative host galaxies.
The fact that all these spectroscopically-examined objects have been 
 confirmed\footnote{Confirmation formally applies only to the optical
 source but, under these circumstances, the probability of a chance 
 positional coincidence is small and it is likely the X-ray and optical
 sources are physically related.} to be background
 sources instead of true ULXs 
 suggests ULXs may not exist beyond the luminous regions of galaxies 
 and hence that there is no association between ULXs and halo objects.

\section{Properties of a ULX Candidate Sample}   

To investigate the merits of this conjecture, 
 a sample of ULX candidates located 
 in the outlying regions of their host galaxies is examined.
The ULX candidate catalog of Colbert \& Ptak (2002) is chosen for this 
 purpose because
 (1) it includes candidates out to twice the optical radius of the 
 target galaxies and (2) \LCG\ (2006) have recently compiled the known optical
 properties of all the ULX candidates in this catalog.
The Colbert \& Ptak (2002) catalog is based on positional coincidences
 between point-like X-ray sources in \ros/HRI images and galaxies listed
 in the  Third Reference Catalogue of Bright Galaxies 
 (RC3; de~Vaucouleurs \etal\ 1991).
The ULX catalog tabulates celestial coordinates and X-ray luminosity estimates
 for 87 ULX candidates detected in 54 galaxy fields (see 
 Colbert \& Ptak 2002; Ptak \& Colbert 2004 for details).

The optical extent of a galaxy is here defined to be the ellipse enclosing
 the isophote at surface brightness 25 B mag arcsec$^{-2}$. 
The major axis diameter (\DTF), major-to-minor axis ratio, and position angle
 of the ellipses are taken from the RC3.
The deprojected angular distance of each ULX candidate 
 from the center of the host galaxy
 can be expressed as the fraction, \fdtf,
 of the angular distance to the \DTF\ isophote along 
 the radius from the galaxy center to the ULX candidate.
Sources outside the optical extent of the host galaxies are defined as
 those with \fdtf$>$1.
Forty-one of the 87 ULX candidates listed in the Colbert \& Ptak (2002)
 catalog have \fdtf$>$1.
Note that the \DTF\ isophote is a reasonable demarcation, on average, between
 the luminous bulge-disk regions of a galaxy and its dark halo while
the specific division at \fdtf$=$1 (and the approximation of the \DTF\ isophote
 as an ellipse) is merely a convenient expedient 
 that could be varied somewhat without altering the results to follow.

A histogram of the distribution of \fdtf\ for all 87 
 ULX candidates is displayed in Figure~1. 
The ordinate is the number, $N($\fdtf$)$, 
 of ULX candidates per unit \fdtf\ area.
Since the angular size, \DTF, varies from galaxy to galaxy, 
 the physical scale represented by unit \fdtf\ area also varies 
 among the ensemble of sources.
However, it does correctly weight each bin such that 
 $N($\fdtf$)$ is independent of \fdtf\ for a uniform spatial
 distribution of sources.
The distribution of ULX candidates in this space is flat in the range
 1$<$\fdtf $<$2 according to Figure~1.
For \fdtf$<$1, the distribution of ULX candidates 
 increases inward to a maximum at \fdtf$=$0.
This is the same trend reported previously for \fdtf$\le$1 ULX candidates
 (cf. Figure~12 of Swartz \etal\ 2004; Figure~16 of Liu \etal\ 2006;
 see also Irwin \etal\ 2004) and roughly 
 follows the distribution of optical emission from the host galaxies.
Beyond \fdtf$=$2, the distribution declines because of incomplete 
 sampling: Colbert \& Ptak (2002)
 considered sources within a {\sl circle} of radius $r=$\DTF\ 
 about the galaxy centers so \fdtf$>$2 occurs only in a subset of azimuthal 
 angles symmetric about the \DTF\ ellipse's minor axis (and favors
 contributions from high-inclination spiral galaxies with
 highly eccentric \DTF\ ellipses).

\begin{center}
\includegraphics[angle=-90,width=0.47\textwidth]{f1.eps} 
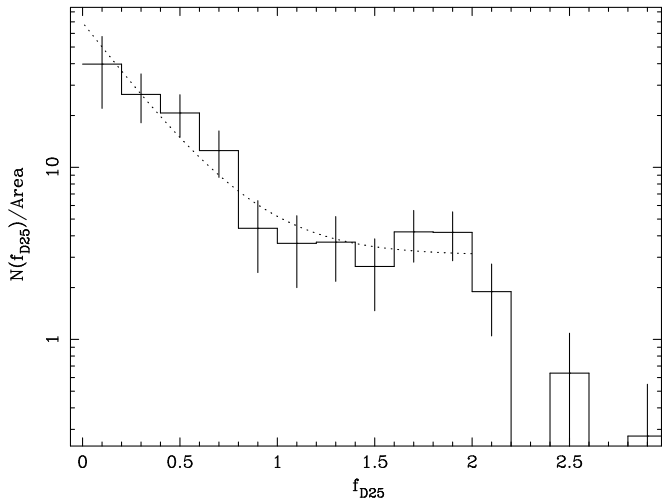
\figcaption{Spatial distribution of 87 ULX candidates from the 
catalog of Colbert \& Ptak (2002). The abscissa is the distance 
of the ULX candidate from the center of its host galaxy in units
of the \DTF\ isophotal diameter and the ordinate is the number
of ULX candidates per unit \fdtf\ area. The dotted curve is the 
best-fit constant plus exponential to the data on the range 
0$<$\fdtf$<$2.
}
\end{center}

A more quantitative statement can be made by fitting
 a constant plus an exponential function to the data on the range 0$<$\fdtf$<$2.
Integrating the best-fit function, $N($\fdtf$)=3.1+66.3\exp(-$\fdtf$/0.3)$
 ($\chi^2=4.7$ for 7 dof), over the range 1$<$\fdtf $<$2 gives a contribution
 of 29.2 sources from the uniform distribution 
 (represented by the constant term)
 and 4.6 sources from the exponential distribution.
This implies 86\% of the ULX candidates beyond \fdtf$=$1 are 
 part of the uniform spatial distribution of background sources.
(There are a total of 33 sources in the sample on this range and another 8 with
 \fdtf$>$2.)
Within \fdtf$=$1, the constant term contributes 9.7 sources and the 
 exponential 30.0 sources implying only 24\% are background.
This is consistent with simply assuming background sources are uniformly
 distributed; from the ratio of areas, 13.7 background sources (30\%) are expected
 within \fdtf$=$1 if all 41 sources beyond \fdtf$=$1 are background.

\LCG\ (2006) have tabulated some optical properties of the ULX candidates
 in the Colbert \& Ptak (2002) catalog. 
Here, only the 41 sources with  \fdtf$>$1 are considered.
Figure~2 displays their
 location in an X-ray-optical flux-flux diagram
 (the X-ray fluxes are also from the \LCG\ tabulation).
X-ray-selected background sources (mostly AGN) detected in deep field 
 (e.g., Brandt \& Hasinger 2005) and in serendipitous wide-field
 (e.g. Green \etal\ 2004) X-ray surveys typically have X-ray-to-optical 
 flux ratios in the range $-1<\log(F_{\rm X}/F_{\rm O})<+1$ with a mean
 around $\log(F_{\rm X}/F_{\rm O})\sim0$. 
Lines depicting $\log(F_{\rm X}/F_{\rm O})=-1$, 0, $+$1 are 
 shown\footnote{The Johnson I or R bands or their
 equivalent are typically used for $F_{\rm O}$ instead of B as quoted in
 \LCG\ (2006) and used here.} in Figure~2. 
The \fdtf$>$1 ULX candidates positionally-coincident with optically-bright 
 counterparts occupy the same range in flux-flux space as do the background 
 X-ray population.

\begin{center}
\includegraphics[angle=-90,width=0.47\textwidth]{f2.eps} 
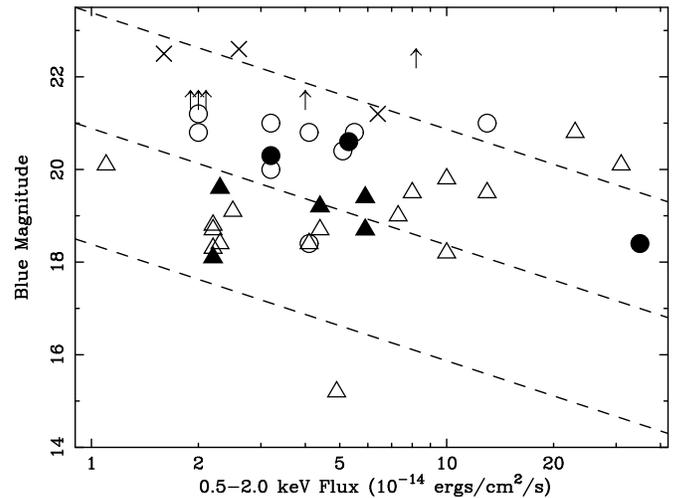
\figcaption{$B$-band magnitude versus 0.5--2.0~keV X-ray flux for
the 41 ULX candidates with \fdtf$>$1. Data from \LCG\ (2006).
Triangles represent objects with spectroscopic redshifts. 
Xs denote sources cataloged in the
Sloan Digital Sky Survey but listed only as upper limits in \LCG.
Arrows denote X-ray sources not detected
in optical data and are located at their $B$ magnitude limit.
Open symbols represent sources in the field of elliptical galaxies and
filled symbols represent those in spiral galaxy fields.
Dashed lines indicate constant $\log(F_{\rm X}/F_{\rm O})=-1$, 0, $+$1.
}
\end{center}

Twenty-one of these sources have had spectroscopic redshifts measured.
These are designated by triangles in Figure~2.
They tend naturally to be the optically-brightest sources.
The redshifts of all 21 of these sources confirm they are background
 objects.
Assuming this is an unbiased sample and taking the distribution of 
 possibilities to be bimodal (they either are or are not background objects), then
 there is a 95\% probability that 90\% or more of
 the population of \fdtf$>$1 ULX candidates are background objects. 
(Note that, since the spectroscopic sample is not randomly
 selected but favors optically-bright objects,
the possibility remains that some or even all of
 the optically-faint ULX candidates are indeed
 an independent population unrelated to the confirmed background sources,
 though they are still distributed uniformly 
 and have $F_{\rm X}/F_{\rm O}$ ratios consistent with background objects.)

\LCG\ (2006) reported potential optical counterparts for all but 8 of the 41
 ULX candidates beyond \fdtf$=$1. 
Three of these 8 are visible in Sloan Digital Sky Survey (SDSS) Data Release 5
 images and designated with Xs in Figure~2
 (using $m_B=g+0.31(g-r)+0.23$ to convert from $g$ and $r$ SDSS magnitudes).
They are all classified by SDSS as galaxies.
Only upper limits based on Digitized Sky Survey (DSS) blue images or SDSS
 are available for the remaining five ULX candidates.
These are marked with arrows at the positions of their X-ray flux and
  limiting optical magnitudes. 
Three of these 5 sources are X-ray faint and fall comfortably below the 
 line $\log(F_{\rm X}/F_{\rm O})=+1$; consistent with background sources.
This leaves two ULX candidates beyond \fdtf$=$1
 that cannot readily be interpreted as background objects
 based on either measured redshifts
 or on the known $F_{\rm X}/F_{\rm O}$ trend among background objects.
The weaker source, at $F_{\rm X}=4\times10^{-14}$~\ergcms, is designated
 IXO~30 in Colbert \& Ptak (2002). 
Bianchi \etal\ (2005) reported an Fe~K$\alpha$ emission
 line in the XMM-{\sl Newton} spectrum of this source and that therefore it
 is likely a foreground Cataclysmic Variable rather than a ULX.
Note that low-mass and even high-mass X-ray binaries are expected to have very
 high $F_{\rm X}/F_{\rm O}$ ratios. 
For example, the optical counterpart to the moderately-luminous ($L_{\rm X}\sim$3$\times$10$^{39}$~\ergl) ULX designated X6 (Fabbiano 1988)
 in the nearby galaxy NGC~3031 is an O9~--~B1 (Swartz \etal\ 2003) or B8~V
 (Liu \etal\ 2002) high-mass star with $B\sim 24$ giving $F_{\rm X}/F_{\rm O}\sim4$.

Figure~2 also shows there 
 are no differences in the distribution of optical or X-ray
 fluxes of sources in the fields near elliptical compared to spiral galaxies.
This is expected if the \fdtf$>$1 sources are background objects 
 unrelated to the putative host galaxies.

Finally, if these \fdtf$>$1 sources are background objects, then their
 number-flux relation should be consistent with that of 
 the resolved cosmic X-ray background. 
To make such a comparison requires knowledge of
 the angular area and the (observation-specific) sensitivity  
 of each X-ray galaxy field in the Colbert \& Ptak (2002) sample 
 including the many observed galaxies that completely lack ULX candidates.
While this level of detail is not readily available, 
 Ptak \& Colbert (2004) have shown that the number of galaxies in the sample
 of Colbert \& Ptak (2002) with ULX candidates above a given luminosity limit
 is identical for both the $r=D_{25}$ search radius used by 
 Colbert \& Ptak (2002) and for the more restrictive $r=0.5D_{25}$ radius once
 the expected background contributions are accounted for.
In other words, the number of ULX candidates per unit area and per unit flux 
 detected beyond \fdtf$\sim$1
 is consistent with the known distribution of background sources.

\section{Discussion}   

For the sample analysed here, the overwhelming majority, and perhaps all, of
 the ULX candidates located beyond the optical extent of their host
 galaxies are unrelated background objects.
This, in itself, is not remarkable since it must be true when \fdtf$\gg$1.
What is of interest is the strong distinction between sources
 inside and outside of \fdtf$\sim$1.
The physical processes that favor production of 
 ULXs within the optical extent of a galaxy must 
 operate rarely, if ever, in galactic halos.  
Since the most relevant X-ray emission mechanism here
 is accretion onto a compact object, 
the lack of halo ULXs means either compact objects are rare 
 or suitable accretion reservoirs are absent in galactic halos.
What are the implications of this for ULX models?

There are two classes of object that have been recognized as potential 
 sources of halo compact objects and in particular as candidate ULXs.
These are the remnants of zero-metallicity Population~III stars 
 and the (possibly massive)
 black holes nurtured in the cores of globular clusters. 

Remnants of the first generation of stars may be massive black holes
 and many could exist in galactic halos
 (Madau \& Rees 2001; Islam \etal\ 2004a).
In order to appear as ULXs, these remnants must be accreting at high
 rates~--~through 
 a thin disk instead of by Bondi-Hoyle accretion 
(King \etal\ 2001)~--~from 
 the interstellar medium or from their own bound relic minihalos.
Their large masses, $\gtrsim$300~\msun, imply they should be strong
 UV/optical sources with $\log(F_{\rm X}/F_{\rm O})\ll1$ 
 (Islam \etal\ 2004b; Volonteri \& Perna 2005). 
There are no ULX candidates in the present sample that meet these
 conditions.

Globular clusters are known to host X-ray binaries; in fact, 
 dynamical interactions in the dense cluster environment
 enhances close binary formation.
ULXs in globular clusters would, in general, have 
 $\log(F_{\rm X}/F_{\rm O})\gtrsim2$ 
 even for the fainter X-ray sources (e.g., Kundu \etal\ 2002; Sarazin \etal\ 2003).
But this is close enough to the optically-faint end of the 
 observed range for \fdtf$>$1 objects 
 that a globular cluster host cannot be completely excluded from the present 
 sample.

The importance of these two ULX scenarios is that Pop~III remnants and 
the dense stellar environments of globular clusters are two potential 
 sites for the elusive intermediate-mass black holes sought to fill
 the gap between stellar-mass and supermassive black holes 
 (van~der~Marel 2004). 
In Pop~III stars, these form from direct collapse of massive stars.
In globular clusters, they may form through the
 gradual accrual of mass over the long lifetime of the cluster, 
 primarily in the form of accreted black holes  (Miller \& Hamilton 2002)
 or gas accretion (Kawakatu \& Umemura 2005), 
 onto a seed black hole of 
 initial mass large enough to prevent ejection from the cluster 
 through recoil.
Alternatively, they may form at the birth of the cluster through runaway
 stellar or proto-stellar mergers (Portegies~Zwart \etal\ 2004). 
(It is unclear if these latter clusters survive to become old globular clusters in 
 galactic halos or if their compact objects would appear as ULXs at this late 
 evolutionary stage). 
This work has shown, whether such objects exist or not, 
 they rarely, if ever, appear as ULXs.

\acknowledgements

I thank the participants of the informal X-ray Aficionados meetings
for motivating discussions and keen insights; 
in particular S. O'Dell, M. Finger, and A. Tennant.
I also thank R. Soria, A. Tennant, and the anonymous referee
 for helpful comments on the manuscript.


\begin{thebibliography}{}

\bibitem[]{1388}
 Bianchi S., Miniutti, G., Fabian, A. C., Iwasawa, K. 2005, MNRAS, 360, 380
\bibitem[]{1388}
 Brandt, W. N., \& Hasinger, G. 2005, ARA\&A, 43, 827
\bibitem[]{1391}
 Colbert, E. J. M., \& Ptak, A. F. 2002, ApJS, 143, 25
\bibitem[]{1405}
 de Vaucouleurs G., de Vaucouleurs A., Corwin Jr. H. G., Buta R. J.,
     Paturel G., \& Fouque P. 1991, Third Reference Catalogue of
     Bright Galaxies (NewYork: Springer-Verlag)
\bibitem[]{1391}
 Fabbiano, G. 1988, ApJ, 325, 544
\bibitem[]{1391}
 Fabbiano, G. 2006, ARA\&A, 44, 323
\bibitem[]{1391}
 Green, P. J., \etal\ 2004, ApJS, 150, 43
\bibitem[]{1391}
 Irwin, J. A., Bregman, J. N., \& Athey, A. E. 2004, ApJ, 601, L143
\bibitem[]{1391}
 Islam, R. R., Taylor, J. E., \& Silk, J. 2004a, MNRAS, 354, 427
\bibitem[]{1391}
 Islam, R. R., Taylor, J. E., \& Silk, J. 2004b, MNRAS, 354, 443
\bibitem[]{347}
 Kawakatu, N., \& Umemura, M. 2005, ApJ, 628, 721
\bibitem[]{347}
 King, A. R., Davies, M. B., Ward, M. J., Fabbiano, G., \& Elvis, M.
 2001, ApJ, 552, L109
\bibitem[]{1483}
 Kundu, A., Maccarone, T. J., \& Zepf, S. E. 2002, ApJ, 574, L5
\bibitem[]{1388}
 L\'{o}pez-Corredoira, M. \& Guti\'{e}rrez, C. M. 2006, A\&A, 454, 77
\bibitem[54]{54_a}
 Liu, J. -F., Bregman, J. N. \& Seitzer, P. 2002, ApJ, 580, L31
\bibitem[54]{54_a}
 Liu, J. -F., Bregman, J. N. \& Irwin, J. 2006, ApJ, 642, 171
\bibitem[54]{54_a}
 Miller, M. C., \& Hamilton, D. P. 2002, MNRAS, 330, 232
\bibitem[54]{54_a}
 Madau, P., \& Rees, M. J. 2001, ApJ, 551, L27
\bibitem[54]{54_a}
 Portegies Zwart, S. F., Baumgardt, H., Hut, P., Makino, J., McMillan, S. L. W.
 2004, Nature, 428, 724
\bibitem[]{1391}
 Ptak, A. \& Colbert, E. 2004, ApJ, 606, 291
\bibitem[]{1391}
 Sarazin, C., Kundu, A., Irwin, J. A., Sivakoff, G. R., Blanton, E. L., Randall, S. W. 2003, ApJ, 595, 743
\bibitem[]{1391}
 Swartz, D. A., Ghosh, K. K., McCollough, M. L., Pannuti, T. G., Tennant, A. F., \& Wu, K. 2003, ApJS, 144, 213
\bibitem[]{1391}
 Swartz, D. A., Ghosh, K. K., Tennant, A. F., \& Wu, K. 2004, ApJS, 154, 519
\bibitem[]{1391}
 van~der~Marel, R.P. 2004, in Coevolution of Black Holes and Galaxies, ed. L.C. Ho
(Cambridge:CUP), 37
\bibitem[]{1391}
 Volonteri, M. \& Perna, R. 2005, MNRAS, 358, 913

\end{thebibliography}
\end{document}